\documentclass[aps,prb,showpacs,twocolumn]{revtex4}
\usepackage{float,graphicx}
\usepackage{psfig}

\usepackage{dcolumn}
\newcommand{\bsim}{\mbox{\raisebox{-0.1cm}{$\;
\stackrel{\textstyle>}{\sim}\;$}}}
\newcommand{\lsim}{\mbox{\raisebox{-0.1cm}{$\;
\stackrel{\textstyle<}{\sim}\;$}}}

\begin{document}

\title{Transport properties
in correlated systems: an analytical model}

\author{F. Rizzo,$^1$
E. Cappelluti$^{2,1}$ and L. Pietronero$^{1,2}$} 

\affiliation{$^1$Dipartimento di Fisica, Universit\`a di Roma ``La Sapienza'', 
P.le A. Moro 2, 00185 Roma, Italy}

\affiliation{$^2$Istituto dei Sistemi Complessi, CNR-INFM,
v. dei Taurini 19, 00185 Roma, Italy} 

\date{\today}

\begin{abstract}
Several studies have so far investigated
transport properties of strongly correlated systems.
Interesting features of these materials
are the lack of resistivity saturation well beyond the
Mott-Ioffe-Regel limit and the scaling of the
resistivity with the hole density in underdoped cuprates.
Due to the strongly correlated nature of these materials,
mainly numerical techniques have been employed. A key role
in this regards is thought to be played by the continuous transfer
of spectral weight from coherent to incoherent states.
In this paper we employ a simple analytical expression
for the electronic Green's function to evaluate both quasi-particle
and transport properties in correlated systems.
Our analytical approach permits to enlighten the specific role
of the spectral transfer due to the correlation
on different features. In particular we investigate the dependence
of both quasi-particle and transport scattering rate on the correlation
degree and the criterion for resistivity saturation.
\end{abstract}
\maketitle

\section{Introduction}

The study of transport properties in high temperature superconductors
represents a still open issue.
Among other debated anomalous features, as the linear behavior of the
resistivity $\rho(T)$ as a function of the temperature
for optimally doped compounds, the resistivity saturation presents
interesting analogies between
the cuprates and the alkali-doped fullerenes.
The concept of resistivity saturation derives from a 
semi-classical picture. 
The basilar argument is that the electronic mean free path
$l$ can not be lower than the inter-atomic distance $a$.
In a Drude-like picture this simple consideration
implies that the conductivity, which is proportional to
the mean free path $l$, cannot be smaller than
a limiting value $\sigma_{\rm sat} = \sigma(l=a)$, and
a saturating value for the resistivity
$\rho_{\rm sat} = 1/\sigma_{\rm sat}$.\cite{Ioffe}
This condition, usually known as the Ioffe-Regel criterion,
is often connected with the Mott limit associated with
the disorder induced metal-insulator transition $k_{\rm F} l \sim
1$.\cite{Mott}

From the experimental point of view, the mostly known materials
displacing resistivity saturation are the A15 compounds, where the
Ioffe-Regel limit is rapidly achieved in the physical range
of temperature due to the strong electron-phonon coupling.
Coherently with this picture, A15 materials
show a pronounced deviation of $\rho(T)$ from the
linear behavior at high temperature when
the mean free path is expected
to become comparable with the inter-atomic distance.\cite{Fisk} 
However, both in cuprates and in alkali-doped fullerenes the
theoretical prediction of the resistivity saturation
at the Ioffe-Regel limit seems to fail.\cite{Ando,Hebard}
Experimental measurements in lightly doped cuprates show a resistivity greater
than the saturation value $\rho_{\rm sat}$ estimated by
the simple quasi-particle picture,\cite{Ando,Takagi,Takenaka}
and also transport measurements in the
alkali-doped compounds would predict at high temperatures
a mean free path smaller than the intermolecular distance.\cite{Hebard}

Although these two classes of materials present significant differences,
an interesting common trait which is shared by both families
is the relevant role of a strong electronic correlation.
The presence of strong correlation effects questions the simple
Drude-like picture based on the quasi-particle concept, and
the above criterion for the resistivity saturation is needed to
be revised in these materials.
On the theoretical ground, 
the lack of resistivity saturation in cuprates and
alkali-doped fullerenes has been recently investigated in details
in Refs. 
\onlinecite{Millis,Gunnarsson_N,Gunnarsson_PRB,Gunnarsson_EPL,Gunnarsson_RPM}
by means of
the numerical Quantum Monte Carlo and dynamical mean-field theory techniques.
A simple analytical criterion, alternative to the Ioffe-Regel one,
has been in addition proposed.
In this alternative framework the resistivity saturation is achieved
when the electron scattering rate $\Gamma$, inversely proportional
to the electron lifetime $\tau$, becomes comparable with the
electron bandwidth $W$. In this regime the quasi-particle concept
is clearly meaningless, the Drude peak is lost and the optical
conductivity $\sigma(\omega)$ is almost structureless
up to energies $\omega \le W$.
By considering a general sum-rule (SR)\cite{Maldague} and an
opportune model for the optical conductivity $\sigma(\omega)$
Gunnarsson and co-workers \cite{Gunnarsson_PRB,Gunnarsson_RPM}
estimated thus 
\begin{equation}
\rho_{\rm sat} \propto W/ T_{\rm K}, 
\label{gunmodel}
\end{equation}
where $T_{\rm K}$ is the electron kinetic energy.
It is interesting to note that, unlike the Ioffe-Regel
criterion which involves the comparison between {\em length}
scales ($l$ and $a$), Eq. (\ref{gunmodel})
relates the resistivity saturation value to the {\em energy}
scales $W$, $T_{\rm K}$.
Note also that Eq. (\ref{gunmodel}) is quite general
and specific material details, as
the source of the electron scattering, are hidden in the evaluation
of the electron quantities $W$, $T_{\rm K}$.
In cuprates for instance, due to the strong
electronic correlation, the amount of charge carriers, and hence
of the kinetic energy, is expected to be roughly proportional
to the hole doping $\delta$, namely
$T_{\rm K} \sim T_{\rm K}^0 \delta(1-\delta)$, where
$T_{\rm K}^0$ is the kinetic energy of the uncorrelated system.
The limiting value of the resistivity saturation is thus expected
to be significantly doping dependent $\rho_{\rm sat}
\sim \rho_{\rm sat}^0/ \delta(1-\delta)$,\cite{Gunnarsson_EPL}
and it could in
particular account for the lack of saturation
in the physical range of temperature
of the low doping regime.

Aim of this paper is to investigate in a more analytical way the
effect of the electronic correlation on the transport properties
and on the saturation phenomenon.
We note that both the Ioffe-Regel criterion ($l \sim a$)
and the simple model introduced by Gunnarsson and coauthors
($\Gamma \sim W/2$) do not rely on the microscopic nature
on the electron scattering mechanism (impurities, phonons) or
on the specific temperature dependence of it, but they are only
related to the absolute magnitude of the scattering channels,
parametrized by the quantity $\Gamma$.
In this perspective, in the following we consider
an impurity scattering mechanism
where the electrons are elastically scattered by dilute paramagnetic
impurities. Although the temperature dependence of the resistivity
in this case is of course disregarded, this assumption is however
sufficient to investigate the phenomenon of the
resistivity saturation which, as mentioned above,
depends only on the absolute value of $\rho$ more than on its temperature
dependence. In our context the impurity scattering rate $\Gamma_0$
plays thus the role of ruling the intensity of the
quasi-particle scattering, in the same way as the temperature
acts within the electron-phonon framework.
At a first level of approximation, thus, our analysis
as function of $\Gamma_0$ can shed a useful light
on the temperature dependence in other scattering mechanisms
once the dependence of $\Gamma_0$ on the temperature is known
($\Gamma_0 \sim$ const. for impurities, $\Gamma_0 \sim T^2$
for electron-electron scattering, 
$\Gamma_0 \sim T^3$ for the
electron-phonon one at low temperatures).
We introduce also an approximate expression for the one-particle
Green's function
to take into account, in a self-consistent way,
the interplay between the electronic correlation and the
impurity scattering. This simple model permits to evaluate in an analytical
way the effects of the electronic correlation on one-particle
quantities as the quasi-particle scattering rate $\Gamma$, the
electron lifetime $\tau$ and the kinetic energy $T_{\rm K}$. 
We evaluate also the
resistivity by using the Kubo's formula, obtaining the behavior of the
resistivity as a function of the correlation degree.
We qualitatively confirm the results of
Refs. \onlinecite{Gunnarsson_EPL,Gunnarsson_RPM}
and we predict a significant difference,
in the low doping highly correlated regime,
between the quasi-particle lifetime as extracted
for instance from angle-resolved photoemission spectroscopy and
the apparent transport lifetime as extracted from resistivity
measurements by using a simple Drude-like formula.

\section{The model}
\label{sec:model}

A paradigmatic model to discuss the strong electronic correlation
effects in solid state physics is the Hubbard model
where propagating electrons described by Bloch-like states
strongly interact each other through a local repulsion.
In the case of a single 
non degenerate band
the Hubbard Hamiltonian reads thus:
\begin{equation}
H=\sum_{ij\sigma}[t_{ij}-\mu\delta_{ij}]c_{i\sigma}^\dagger
c_{j\sigma}
+\frac{U}{2}\sum_{i\sigma\sigma'}n_{i\sigma}n_{i\sigma'}
\label{H1}
\end{equation}
where $n_{i\sigma}=c_{i\sigma}^\dagger c_{i\sigma}$,
$t_{ij}$ represents 
tight-binding hopping elements and $U$ describes an on site Coulomb repulsion.
The chemical potential $\mu$ rules the electronic band filling $n$.
In the most interesting case close at half-filling ($n=1$),
the complex physics of the Hubbard model can be parametrized
in terms of two energy scales: the kinetic energy, which scales
with the electronic bandwidth $W$, and the Hubbard repulsion $U$.
The behavior of the system 
depends strictly on the ratio $U/W$, where the $U/W \ll 1$ regime
describes a weakly correlated system with strong
itinerant quasi-particle character, while in the $U/W \gg 1$ limit
the electronic states are almost totally localized.

A standard way to describe electronic properties in interacting systems
is by means of the Green's function formalism in the
context of the Quantum Field Theory.
Single-particle properties are thus taken into account by
Green's function $G({\bf p},\omega)$ which describes
the propagation of a one-particle electronic excitation with
momentum ${\bf p}$ and energy $\omega$.
From a generic point of view,
the interacting Green's function $G({\bf p},\omega)$
is usually described in terms
of a {\em coherent} and an {\em incoherent} part:\cite{Abrikosov}
\begin{equation}
G({\bf p},\omega)=G_{\rm coh}({\bf p},\omega) +
G_{\rm inc}({\bf p},\omega),
\label{Green}
\end{equation}
where $G_{\rm coh}({\bf p},\omega)$
describes quasi-particle 
Bloch-like electrons for which ${\bf p}$ is a good quantum number, 
and $G_{\rm inc}({\bf p},\omega)$
corresponds to the incoherent background with a weak
dependence on the electronic momentum: $G_{\rm inc}({\bf p},\omega)
\simeq G_{\rm inc}(\omega)$.

Different analytical approaches have been employed to deal with
the Hubbard model,
according whether the main interest is paid on
the coherent or the incoherent part in the strong correlated
regime.
An example of the first case is
Gutzwiller method,\cite{Gutzwiller} reproduced by the mean-field
solution of the slave-boson techniques, which
describes a coherent quasi-particle state
with reduced spectral weight $Z$ ($Z < 1$) and
reduced bandwidth $W^{\rm eff} = Z W$.   
In the half-filling case and for $U$ larger than a critical value
$U \ge U_c$, $Z$ vanishes describing a metal-insulator
Brinkmann-Rice transition.\cite{Rice}
From the opposite point of view, the Hubbard I (Hub-I)
approximation,\cite{Hubbard}
which is exact in the atomic limit, is mainly aimed at a schematic
representation
of the localized states, described by an upper
and a lower Hubbard band spaced by an energy gap of width $U$.
A comprehensive description of the system should account in a mixed way
for both these features, as it is confirmed by numerical calculations
based on the dynamical mean-field theory.\cite{Kotliar}
The actual predominance of itinerant quasi-particle states
or of almost localized
incoherent excitations is ruled by their respective
spectral weights, $Z_{\rm coh}$ and $Z_{\rm inc}=1-Z_{\rm coh}$,
which depend on the microscopic parameters
$U$ and $n$. Physical properties are expected to be strongly dependent
of the amount of the coherent and incoherent spectral weight.

In this paper we introduce a simple analytical model to describe
the transfer of spectral weight from coherent to incoherent states,
and its effects on transport properties.
In explicit terms,we approximate
the (unknown) coherent part $G_{\rm coh}({\bf p},\omega)$ with the Gutzwiller
solution and the incoherent term with the Hubbard I solution.
The retarded Green's function reads thus:\cite{notegreen}
\begin{eqnarray}
G_{\rm coh}({\bf p},\omega)&=&\frac{Z}{\omega-Z\epsilon_{\bf p}+\mu},
\label{Greenc}
\\
G_{\rm inc}(\omega)&=&\frac{1-Z}{N_s}
\sum_{\bf p}\left[
\frac{n/2}
{\omega-\epsilon_{\bf p} n/2+\mu+U/2}\right]
\nonumber
\\
&&\left.
+\frac{1-n/2}
{\omega-(1-n/2)\epsilon_{\bf p}+\mu-U/2}\right],
\label{Green1}
\end{eqnarray}
where $Z$ is the quasi-particle spectral weight
evaluated by the Gutzwiller solution for generic $U$ and $n$,\cite{Lilia}
$N_s$ is the total number of sites, and where the factor $(1-Z)$ has
explicitly introduced in the incoherent term in order to preserve
the conservation of the total spectral weight.
In addition,
due to the quasi-localized picture of the electronic states in the
Hubbard bands, we have also approximated the incoherent Green's function
with its local contribution, so that
$G_{\rm inc}(\omega) \simeq (1-Z)/N_s
\sum_{\bf p} G_{\rm Hub-I}({\bf p},\omega)$.
We want to stress that
the model introduced in Eqs. (\ref{Green})-(\ref{Green1})
is not meant to be exhaustive of
all the complex phenomenology of correlated systems.
However, it has the advantage to
account in the simplest and analytical way for the transfer of
spectral weight, here ruled by the parameter $Z(U,n)$,
from coherent to incoherent
states by increasing the rate of the electronic correlation.
This simple model was recently employed to study the reduction
of the screening properties in strongly correlated systems,
leading to a predominance of forward scattering,
and its interplay with the nonadiabatic superconducting pairing.\cite{Lilia}
Simple arguments to understand the reduction of the screening properties
in correlated systems come from the fact that the metallic screening
of charge fluctuations requires the momentum ${\bf p}$ to be a good
quantum number and it is mainly related to the coherent states.
As we are going to see, similar argumentations hold true also
for transport properties.

In the following we are going to employ the model described
by Eqs. (\ref{Green})-(\ref{Green1})
to investigate correlation effects on  the transport properties
induced by impurity scattering. As a first step toward this aim we need
therefore to generalize Eqs. (\ref{Green})-(\ref{Green1})
in the presence of impurities.
A phenomenological way to take into account scattering by nonmagnetic
impurities is by introducing an impurity ``self-energy'' term
$\Sigma_{\rm imp}(\omega)$ which renormalizes the bare 
electronic frequency $\omega$:
\begin{equation}
\omega \rightarrow \omega - \Sigma_{\rm imp}(\omega).
\label{impfreq}
\end{equation}
The explicit expression of the electron Green's function
in the presence of impurity reads thus:
\begin{equation}
G_{\rm coh}({\bf p},\omega)=
\frac{Z}{\omega-Z\epsilon_{\bf p}+\mu 
-\Sigma_{\rm imp}(\omega)},
\label{Green2c}
\end{equation}
\begin{eqnarray}
G_{\rm inc}(\omega)&=&\frac{1-Z}{N_s}
\nonumber\\
&&\times \sum_{\bf p}\left[
\frac{n/2}
{\omega-\epsilon_{\bf p}n/2+\mu+U/2-\Sigma_{\rm imp}(\omega)}\right.
\nonumber\\
&&\left. 
+\frac{1-n/2}
{\omega-(1-n/2)\epsilon_{\bf p}+\mu-U/2-\Sigma_{\rm imp}(\omega)}\right].
\label{Green2i}
\end{eqnarray}
Note that, generally speaking, taking into account the impurity scattering
through a ``self-energy'' term as done in
Eqs. (\ref{impfreq})-(\ref{Green2i}) is not formally correct since
it assumes: $i$) that each contribution of the Green's function
(coherent part, upper and lower Hubbard band) conserves the same structure
as in the absence of impurities
with a simple replacing
$\omega \rightarrow \omega - \Sigma_{\rm imp}(\omega)$; $ii$) that
the  ``self-energy'' term is the same for each of those contribution.
In this way we are neglecting then the interferences between impurity
and Coulomb scattering and between coherent and incoherent terms
which can in principle take place.
In our approach we assume Eq. (\ref{impfreq}) thus to be valid
{\em as an approximation}.
We note however that Eqs. (\ref{Green2c})-(\ref{Green2i}),
although approximate, are expected to work well
both in the uncorrelated limit $U=0$, where the impurity self-energy
$\Sigma_{\rm imp}(\omega)$ is related to the resummation of the
$T$-matrix of the impurity scattering, and in highly correlated
limit $W = 0$ where no electronic hopping is allowed, the electrons
are localized and impurity scattering does not give rise
to any decay processes ($\Sigma_{\rm imp}''(\omega)=0$).

\section{One-electron self-energy and quasi-particle lifetime}
\label{sec:qpt}

In the previous section we have introduced an analytical
model for the electron Green's in the presence of both
electronic correlation and impurity scattering.
An important feature of this model is to account in a
simple way for the transfer of spectral weight between itinerant
and localized states as function of the degree of correlation.
In this section we employ the model to estimate the effect of the
electronic correlation on the finite quasi-particle lifetime due to
the impurity scattering.
In order to evaluate the impurity self-energy we assume the
$T$-matrix resummation,\cite{Mahan} which is exact for uncorrelated systems
in the dilute impurity regime ($n_{\rm imp}\ll 1$), to be valid
even in the presence of electronic correlation, namely:
\begin{equation} 
\Sigma_{\rm imp}(\omega)=
\frac{n_{\rm imp} V_{\rm imp}}
{1- V_{\rm imp} G_{\rm loc}(\omega)},
\label{selfimp}
\end{equation}
where $ G_{\rm loc}(\omega)$ is the 
local Green's function:
\begin{equation}
G_{\rm loc}(\omega)=
\frac{1}{N_s}\sum_{\bf p}G({\bf p},\omega)
=G_{\rm loc}'(\omega)+iG_{\rm loc}''(\omega),
\label{G_loc}
\end{equation}
and $V_{\rm imp}$ the electron-impurity matrix element, here
assumed to be constant.
Note that, unlike the case of retarded interactions,
Eq. (\ref{selfimp}) relates the self-energy evaluated
at the frequency $\omega$, $\Sigma_{\rm imp}(\omega)$,
only to the Green's function evaluated at the {\em same}
frequency, $G_{\rm loc}(\omega)$.
This gives a significant advantage because,
since we are mainly interested in transport and electronic
properties close to the Fermi level, it is sufficient
to evaluate, in a self-consistent way,
the zero frequency limit of the impurity self-energy
$\Sigma_{\rm imp}(\omega) \simeq \Sigma_{\rm imp}(\omega=0)$,
without involving the full frequency structure
of $\Sigma_{\rm imp}(\omega)$.
In this perspective we approximate
the impurity self-energy with only two parameters,
$\Delta$ and $\Gamma$,
representing respectively the real and imaginary part
of the impurity self-energy at the Fermi level,
$\Sigma_{\rm imp}=\Delta-i\Gamma$.
The impurity parameters $\Delta$ and $\Gamma$
will be thus evaluated in a self-consistent way
as function of 
the degree of electronic correlation and of microscopic
quantities as $n$, $U$ and the coherent spectral weight $Z(n,U)$.
We have:
\begin{eqnarray}
\Delta &=& \frac{n_{\rm imp} V_{\rm imp}[1-V_{\rm imp}G_{\rm loc}']}
{[1-V_{\rm imp}G_{\rm loc}']^2+[ V_{\rm imp}
G_{\rm loc}'']^2},
\label{gammadelta}
\\
\Gamma &=&-\frac{n_{\rm imp} V_{\rm imp}^2
  G_{\rm loc}''}
{[1-V_{\rm imp}G_{\rm loc}']^2+[ V_{\rm imp}
G_{\rm loc}'']^2}.
\label{gammaz}
\end{eqnarray}
In the absence of electronic correlation, for
an infinite bandwidth system, $G_{\rm loc}'=0$
and $G_{\rm loc}''=-\pi N(0)$,
and we obtain the usual relations,
$\Delta(U=0,W\rightarrow \infty)= \Delta_0/
[1+(\pi N(0)V_{\rm imp})^2]$,
$\Gamma(U=0,W\rightarrow \infty)=
\Gamma_0/
[1+(\pi N(0)V_{\rm imp})^2]$, where $N(0)$ is electron
density of states (DOS) at the Fermi level and where
$\Delta_0=n_{\rm imp} V_{\rm imp}$,
$\Gamma_0=n_{\rm imp} \pi N(0)V_{\rm imp}^2$, are respectively
the weak-coupling mean impurity potential and
the weak-coupling impurity scattering rate.

In the following, in order to provide an analytical evaluation
of the correlation effects on the impurity scattering, we employ
a constant DOS model with
$N(\epsilon) = N(0)$
for $|\epsilon| \le W/2$, for which case
the critical Hubbard energy is $U_c=2W$.
This constant DOS model is here meant to be representative
of three-dimensional systems with structureless DOS (as the fullerenes)
and also of two-dimensional systems as the cuprates
provided the chemical potential is not too close to the
logarithmic Van Hove singularity.
Several energy scales can be identified in the system, among which
the Hubbard repulsion $U$,
the bare electron bandwidth $W$, the ``effective'' bandwidth of the
coherent states $W^{\rm eff} = Z W$, the bare impurity scattering rate
$\Gamma_0 = \pi N(0) n_{\rm imp} V_{\rm imp}^2$, and
the effective impurity scattering rate $\Gamma$ ($\le \Gamma_0$)
which will be
introduced in the following.
Aim of this paper is to focus on the effects on the impurity scattering
of the spectral weight transfer induced by the electronic
correlation.
In this perspective we consider a system where in the absence
of correlation the impurity scattering rate 
is sufficiently small compared to the electronic bandwidth,
$\Gamma_0 \ll W$, so that finite bandwidth effects can be neglected.
It should be noted however that in the presence of electronic correlation
a new electronic energy scale appears, namely
the coherent quasi-particle bandwidth $W^{\rm eff} = ZW$, in addition to the
energy scales $W$ and $U$ which still characterize the Hubbard subbands.
Close the metal-insulator transition the shrinking of the
coherent bandwidth can be so operative that
$\Gamma \sim W^{\rm eff}/2 = ZW/2$.
As we are going to see, in order to preserve physical results,
it is thus important the large bandwidth limit 
$\Gamma_0 \ll W$
to be not confused with $\Gamma / W^{\rm eff} \ll 1$.

The impurity self-energy can now be analytically computed by using
the constant DOS model.
Generally speaking, both the coherent and incoherent parts of
the one-particle Green's function will contribute to the
impurity self-energy (see Appendix \ref{appendice} for an explicit
expression of each contribution). As we are going to show, however,
the total impurity self-energy is mainly dominated by 
the only coherent part. From the self-consistent evaluation
of Eqs. (\ref{gammadelta})-(\ref{gammaz}) we can also
determine the corresponding
quasi-particle lifetime $\tau$ through the relation $\tau=\hbar/2\Gamma$,
while the effective impurity potential $\Delta$ is
usually disregarded since it gives just a shift of the chemical potential.

\begin{figure}[t]
\centerline
{\psfig{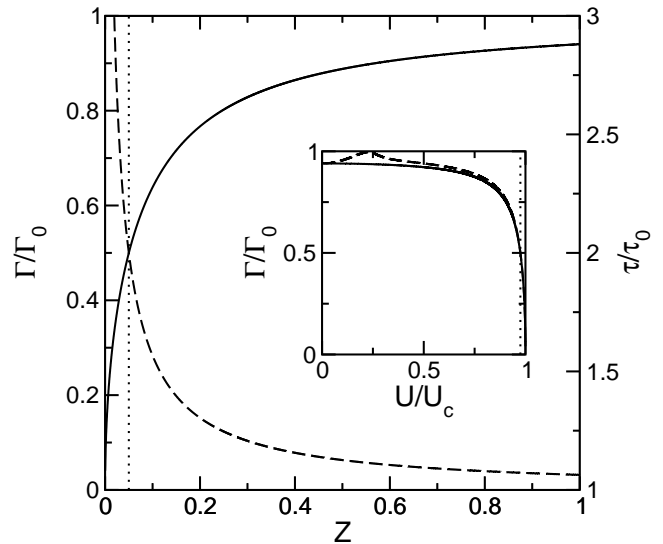}}
\caption{Impurity scattering rate $\Gamma/\Gamma_0$ (solid line,
left side scale)
and quasi-particle lifetime $\tau/\tau_0$ (dashed line, right side
scale) as function of the quasi-particle spectral weight $Z$ for
$\mu=0$, $\Gamma_0/W=0.05$ and $ \pi N(0)V_{\rm imp} = \Gamma_0/W$.
Inset: same quantity $\Gamma/\Gamma_0$ (solid line) plotted as
function of the reduced Hubbard repulsion $U/U_c$ and compared
to the scattering rate where the incoherent term are
fully included (dashed line) as discussed in the text.
The vertical dotted lines mark the condition $\Gamma=W^{\rm eff}/2$.}
\label{tau}
\end{figure}

In Fig. \ref{tau} we plot the behavior of the impurity scattering rate
$\Gamma/\Gamma_0$ (solid line, left side scale)
as well as the quasi-particle scattering time
$\tau/\tau_0$ (dashed line, right side scale)
as function of the amount of electronic
correlation parametrized by the quasi-particle spectral weight $Z$.
As the most representative case of the strongly correlated regime
we consider a half-filling case $\mu=0$
where $Z(U)$ is univocally determined by the Hubbard repulsion $U$:
$Z=1$ corresponds thus to the uncorrelated limit $U=0$ while $Z \rightarrow 0$
describes the Brinkmann-Rice metal-insulator transition for
$U \rightarrow U_c^-$.\cite{Rice} Other microscopic impurity parameters
are set $\Gamma_0/W=0.05$ and $\pi N(0)V_{\rm imp} = 0.1 \Gamma_0/W$.
Since scattering processes are mainly determined by the coherent propagating
electrons, in the evaluation of the quasi-particle scattering rate
$\Gamma$ we have neglected the incoherent contributions of the local
Green's function (Eqs. (\ref{g_{inc}}) and (\ref{f_{inc}})).
The behavior of $\Gamma/\Gamma_0$ as function of $U/U_c$ is also
shown in the inset where we compare the $\Gamma$ evaluated
by taking into account only the coherent part (solid line curve)
with the total scattering rate including the incoherent terms
(dotted line curve).
The small discrepancy between the two curves
points out that the scattering rate is mainly determined by
the coherent part of the one-particle Green's function.
The small hump of $\Gamma$ (dashed line) in the inset is due
to the only incoherent part. It signalizes the range where
the incoherent Green's function have its highest spectral weight
{\em at the Fermi level}, and it roughly corresponds to 
$U \simeq W/2$ where a metal-insulator transition is expected
in the simple Hubbard I model.

Fig. \ref{tau} points out two main regimes: a low correlation regime,
ruled by the parameter $2\Gamma/W^{\rm eff} \lsim 1$, where the
scattering rate $\Gamma$ depends very weakly on the degree of
the electronic correlation; and a strong correlation regime,
characterized by $2\Gamma/W^{\rm eff} \bsim 1$,
where the scattering rate $\Gamma$ is strongly reduced as
$Z \rightarrow 0$.
In this latter regime both the effective bandwidth
$W^{\rm eff}$ of the coherent states and the scattering rate $\Gamma$
vanish for $Z \rightarrow 0$, but with a different
behavior, $W^{\rm eff} \propto Z$ and
$\Gamma \propto \sqrt{Z}$
[$\Gamma(Z)\simeq \sqrt{\Gamma_0ZW/\pi}$].
In this regime the imaginary part of the one-particle self-energy
is thus larger than the coherent bandwidth itself,
the electronic momentum {\bf p} is no more a good quantum number,
and the concept of quasi-particles breaks down.

On the right side scale of Fig. \ref{tau} (dashed line)
we show also the corresponding quasi-particle lifetime $\tau$,
which is simply given by $\tau=\hbar/2\Gamma$.
The vanishing of $\Gamma$ for $Z \rightarrow 0$ corresponds
thus to an infinite lifetime $\tau \rightarrow \infty$.
This is understandable considering that for $Z \rightarrow 0$
the electronic system is completely localized and impurity scattering
cannot trigger any decay processes.
This seems in apparent contradiction with the previous scenario
where we have described this regime as a non quasi-particle one
where scattering rate is larger than the bandwidth itself.
We remind however the quasi-particle properties are ruled
by the {\em ratio} $2\Gamma/W^{\rm eff}$. As we have seen above,
this parameter can be significant larger $2\Gamma/W^{\rm eff} \gg 1$,
pointing out the failure of the quasi-particle concept,
even if $\Gamma \rightarrow 0$ since the effective bandwidth $W^{\rm eff}$
vanishes quicker than $\Gamma$ for $Z \rightarrow 0$.
Note that, in this perspective, the self-consistent
evaluation of the impurity scattering rate $\Gamma$ is fundamental
to recover the correct physical limits for $Z \rightarrow 0$.
If we would evaluate the impurity self-energy from Eq. (\ref{selfimp})
by using Eqs. (\ref{Greenc})-(\ref{Green1}) instead of
Eqs. (\ref{Green2c})-(\ref{Green2i}) we would predict indeed
a finite $\Gamma$ for $Z \rightarrow 0$, since the
limit $\Gamma \ll W^{\rm eff}/2$ is in this case implicitly enforced.

\section{Transport properties}
\label{transport}

In the previous section we have found
how the electronic correlation influences 
quasi-particle properties like the self-energy and the quasi-particle
scattering time. Now we focus on transport properties of
correlated systems.
In particular we are interested on the resistivity $\rho$, or,
equivalently, on the electrical conductivity $\sigma = 1/\rho$
which is evaluated as
\begin{eqnarray}
\sigma=-\lim_{\omega \rightarrow 0}\frac{\mbox{Im}\Pi(\omega)}{\omega},
\label{cond}
\end{eqnarray} 
where $\mbox{Im}\Pi(\omega)$ is the retarded part of
the current-current correlation function,
\begin{eqnarray}
\Pi(\omega)=-\frac{i\hbar}{3 V_{\rm cell}}
\int_{-\infty}^{\infty}dt e^{i\omega t}\theta(t)
\langle[{\bf j} ^{\dagger}(t),{\bf j}(0)]\rangle,
\label{correl.}
\end{eqnarray}
and where ${\bf j}$ is the current operator
${\bf j}= (e/N_s) \sum_{{\bf p},\sigma}{\bf v}_{\bf p}
c_{{\bf p},\sigma}^\dagger c_{{\bf p},\sigma}$, and
$V_{\rm cell}$ the volume of the unit cell.

\begin{figure}[t]
\centerline
{\psfig{figure=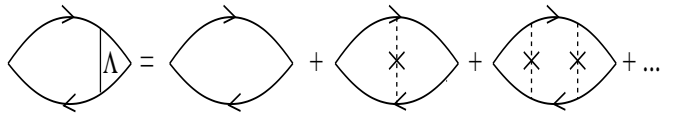,width=9cm,clip=}}
\caption{Diagrammatic representation of the current-current correlation
function. Solid lines represent the one-particle
Green's function in the presence of impurity as given by
Eqs. (\ref{Green2c})-(\ref{Green2i}), dashed line the
interaction between electrons and impurities (crosses).
We neglect here interference of scattering between two or more
impurities, which is unimportant in the limit of small impurities
concentration $n_{\rm imp}\ll 1$.}
\label{bubbles}
\end{figure}

The evaluation of the conductivity requires thus
the evaluation of $\Pi(\omega)$, which is in principle
a two-particle function of the system.
A standard way to relate $\Pi(\omega)$ to the one-particle
Green's function is through a skeleton diagram expansion which
can be obtain by expanding Eq. (\ref{correl.})
as function of the impurity scattering term.
From a diagrammatic point of view, this
expansion corresponds to the infinite sequence of bubble
diagrams in which there are one or more impurity scatterings which link 
the Green's functions on both sides of the bubble (Fig.~\ref{bubbles}).
In a Drude-like theory the current-current response function
$\Pi(\omega)$ is approximated with the only first term of
Fig. \ref{bubbles}, in the so-called
simple bubble approximation.
In this scheme the DC electrical conductivity simply reads:\cite{Mahan}
\begin{equation}
\sigma = \frac{2\pi \hbar e^2}{3N_sV_{\rm cell}}\sum_{\bf p}
|v_{\bf p}|^2
\int d\omega A^2({\bf p},\omega)
\left[-\frac{d n_{\rm F}(\omega)}{d\omega}\right],
\label{convol}
\end{equation}
where $n_{\rm F}(\omega)$ is the Fermi-Dirac distribution and
$A({\bf p},\omega)$ is the electron spectral function
$A({\bf p},\omega)=-(1/\pi)G''({\bf p},\omega)$.
In our model for correlated systems the electronic spectral
function is composed by a coherent and an incoherent term: 
\begin{equation}
A({\bf p},\omega)=A_{\rm coh}({\bf p},\omega)+
A_{\rm inc}(\omega),
\end{equation}
so that the resulting conductivity
 contains three contributions:
\begin{equation}
\sigma = \sigma_{\rm coh}+\sigma_{\rm coh-inc}+\sigma_{\rm inc}.
\label{sigma}
\end{equation}   
The term $\sigma_{\rm coh}$ is associated to coherent
scattering events between states with a well-defined momentum. The
other two contributions are related to the incoherent part of the
electronic Green's function; in particular, $\sigma_{\rm coh-inc}$ takes
into account for the interference between coherent and incoherent
states, while in $\sigma_{\rm inc}$ scattering processes
associated to the excitations between the two Hubbard bands are
considered.

\begin{figure}[t]
\centerline
{\psfig{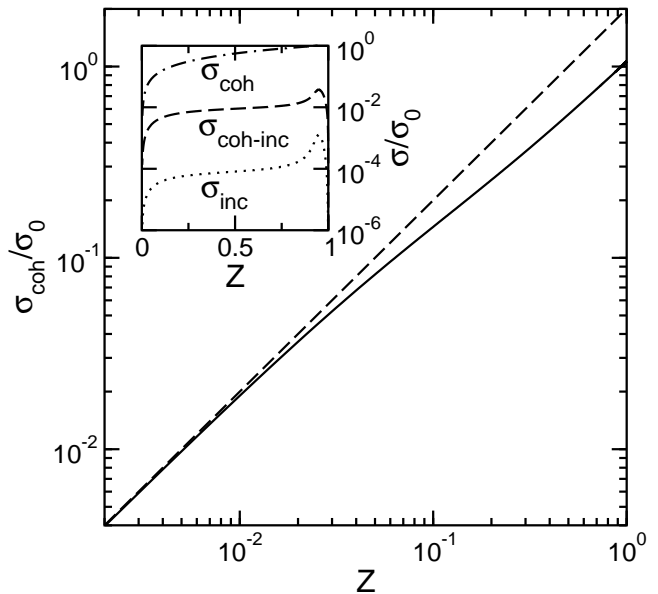}}
\caption{Coherent contribution
to the electrical conductivity $\sigma/\sigma_0$ from the simple bubble 
approximation (solid line) and asymptotic value of the coherent
term in the limit $ZW/2\Gamma \ll 1$ (dashed line) as function of
the spectral weight parameter $Z$. Inset: comparison between the
coherent and incoherent conductivity components.
Microscopic parameters and details as in Fig. \ref{tau}.}
\label{condnov}
\end{figure}

By using the constant DOS model above introduced and
using also a constant electron velocity
$|v_{\bf p}|^2 \simeq |v_{\rm F}|^2$, we obtain:
\begin{eqnarray}
\sigma_{\rm coh}&=&
\sigma_0 \frac{\Gamma_0}{\Gamma}
\frac{Z}{\pi}
\left[I_{\rm coh}(Z)+A(Z)\right],
\label{s_coh}
\\
\sigma_{\rm coh-inc}&=&
\sigma_0\frac{4N(0)\Gamma_0 (1-Z)}{\pi}
I_{\rm coh}(Z)I_{\rm inc}(Z,U),
\\
\sigma_{\rm inc}&=&
\sigma_0 \frac{2N(0)\Gamma_0 (1-Z)^2}{\pi}
I_{\rm inc}^2(Z,U),
\label{s_inc}
\end{eqnarray}
where
\begin{equation}
\sigma_0=\frac{e^2\hbar v_{\rm F}^2 N(0)}{3V_{\rm cell}\Gamma_0}
\label{sig0}
\end{equation}
is the weak scattering ($\Gamma_0 \ll W/2$)
uncorrelated electrical conductivity
and where the explicit expressions of the functions
$I_{\rm coh}$, $I_{\rm inc}$, $A$
are reported in Appendix \ref{appendice}.
The functions $I_{\rm coh}$, $I_{\rm inc}$
are defined to be regular ($\sim$ const.)
for $2\Gamma/W^{\rm eff} \ll 1$ and $A \rightarrow 0$
in the same limit, so that the leading term in the weak scattering
$\Gamma_0 \rightarrow 0$
or dilute impurity $n_{\rm imp} \rightarrow 0$ limits
is the coherent one, $\sigma_{\rm coh} \propto \sigma_0$,
while the other contributions involving incoherent
spectral weight are negligible
[$\sigma_{\rm coh-inc}$, $\sigma_{\rm inc} 
\propto \sigma_0 N(0) \Gamma_0$].
The relevance of each contribution of $\sigma$
in the whole range of correlation is shown in the inset
of Fig. \ref{condnov} which points out that 
the electrical conductivity is mainly dominated by the coherent
terms of the spectral function.

The dependence of only the coherent contribution to
$\sigma$ is also shown in the main panel of Fig. \ref{condnov}
on a double-logarithmic scale in order to
point out the asymptotic behavior of $\sigma$
in both weakly and strongly correlated regimes,
respectively $Z \rightarrow 1$ and $Z \rightarrow 0$.
We find a linear behavior of the electrical conductivity
with $Z$ in both of these regimes.
The weak correlation regime, characterized by
$2\Gamma/W^{\rm eff} \ll 1$, is essentially identical to the
dilute impurity limit, $\Gamma \sim \Gamma_0$ (see Fig. \ref{tau})
and a linear behavior $\sigma_{\rm coh} = Z\sigma_0$
is essentially enforced by the spectral factor $Z$ weighting
the coherent processes.
More delicate is the opposite case of strong correlation
$Z \rightarrow 0$. In this limit, as discussed
in section \ref{sec:qpt}, the impurity scattering rate is
no more the smallest energy scale of the system
since $W^{\rm eff}/2 = ZW/2 \ll \Gamma$, and a
$ZW/2\Gamma \ll 1$ expansion should be employed.
Different energy scales, $\Gamma$, $W^{\rm eff}$ go to zero 
as $Z \rightarrow 0$, as the total effect of the
electronic correlation of the electrical conductivity
has to be determined by the asymptotic behaviors.
In particular, using the asymptotic expressions
$W^{\rm eff} \sim ZW$ and $\Gamma \sim \sqrt{\Gamma_0ZW/\pi}$
valid for $Z \rightarrow 0$, we can check in an analytical way that
in the dilute limit $n_{\rm imp} \propto 2\Gamma_0/W \ll 1$
$\sigma_{\rm coh-inc}/\sigma_{\rm coh} \ \sim \Gamma_0$
and $\sigma_{\rm inc}/\sigma_{\rm coh} \sim \Gamma_0^2$
(we use here also the property $U \rightarrow U_c^- \ge W/2$) 
so that the coherent contribution to the
electrical conductivity $\sigma_{\rm coh}$ is dominant
even in this regime.
The asymptotic behavior of $\sigma_{\rm coh}$ as function
of $Z$ can be determined by the same expansion,
$\sigma_{\rm coh} = 2Z \sigma_0$, resulting
is a linear dependence even in this regime, although with
a different prefactor.
The crossover between these two opposite regimes is shown
in the main panel of Fig. \ref{condnov} (solid line)
where the asymptotic behavior of $\sigma_{\rm coh}$
for $Z \rightarrow 0$ (dashed line) has been also superimposed.

Fig. \ref{condnov} shows that in both weakly and strongly
correlated regimes the conductivity, which is substantially given by
the coherent term, scales with the spectral weight parameter $Z$,
so that the resistivity, $\rho = 1/\sigma$, scales with the inverse of the
quasi-particle spectral weight $Z$.
In physical terms this means that as the degree of the 
electronic correlation is increased, the electronic states
become localized, the conductivity vanishes and the resistivity diverges.

Let us now discuss in which way the present results
affect the analysis of the experimental resistivity
measurements. From the experimental point of view,
a common way to analyze transport properties is by means of
a so-called phenomenological Drude-like model,
where the electrical conductivity is expressed as:\cite{Timusk}
\begin{equation}
\sigma= \frac{e^2 \hbar v_{\rm F}^2 N(0)}{3V_{\rm cell}\Gamma_{\rm tr}},
\end{equation}
where $v_{\rm F}$, $N(0)$ are one-particle properties estimated
by band structure calculations in the absence of electronic
correlation and where $\Gamma_{\rm tr}$ represents an
``effective'' transport
scattering rate.
The same Drude-like model predicts a quasi-particle scattering rate
$\Gamma$ which is usually assumed to be of the same order
of $\Gamma_{\rm tr}$.

If we apply this phenomenological model to our correlated case
we find an {\em effective} Drude-like transport
scattering rate $\Gamma_{\rm tr}$:
\begin{equation}
\Gamma_{\rm tr}= \frac{\pi \Gamma}{Z \left[I_{\rm coh}(Z)+A(Z)\right]},
\label{Gamma_transport}
\end{equation}
where $\Gamma$ is the quasi-particle scattering rate defined
in a self-consistent way by Eq. (\ref{gammaz}).
Eq. (\ref{Gamma_transport}) suggests that
the transport and quasi-particle scattering rates,
extracted in a phenomenological way from the experimental data
within a simple Drude-like model, can significantly
differ each other.

\begin{figure}[t]
\centerline{
\psfig{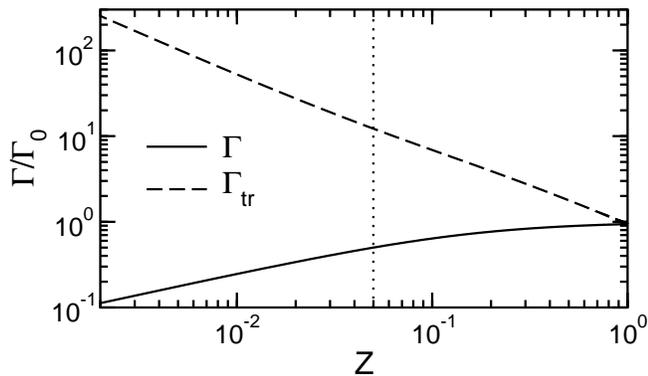}}
\caption{Quasi-particle scattering rate (solid line)
and transport scattering rate (dashed line)
as function of the spectral weight
parameter $Z$ in the half-filling case.
Microscopic parameters and details as in Figs. \ref{tau} and
\ref{condnov}.}
\label{gamma_tr}
\end{figure}
In Fig. \ref{gamma_tr} we compare the behavior of the
quasi-particle $\Gamma$ and of the transport scattering rate $\Gamma_{\rm tr}$
as function of the spectral weight parameter $Z$.
As mentioned above,
$\Gamma_{\rm tr}$ differs significantly from the quasi-particle scattering
rate $\Gamma$ for any finite degree of electronic correlation
and they approach the same value only in the uncorrelated limit
$Z \rightarrow 1$.

\section{Resistivity saturation and correlation effects}
\label{sec-sat}

\begin{figure}[t]
\centerline{
\psfig{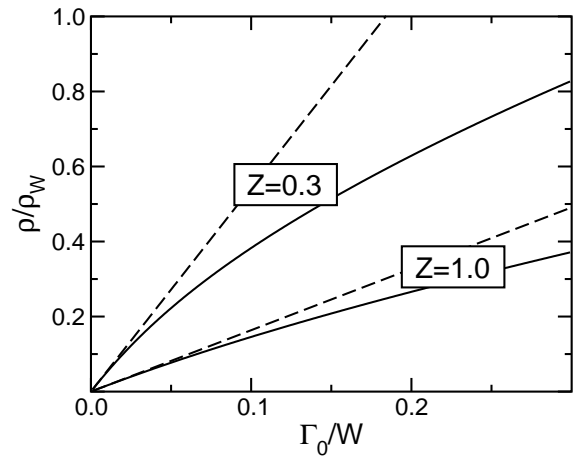}}
\caption{Dependence of the DC resistivity as function of the bare
impurity scattering rate for two values of the amount of
electronic correlation (solid lines). The DC resistivity is plotted in units
of $\rho_W = 3 V_{\rm cell} W / e^2 \hbar v_{\rm F}^2 N(0)$
and $\Gamma_0$ in units of $W$. The dashed lines
represent the linear asymptotic behavior in the weak
$\Gamma_0 \rightarrow 0$ limit.
}
\label{f-satg}
\end{figure}

So far our analysis
has focused on the effects of the electronic correlation
on the quasi-particle and transport properties
in a regime where the coherent part of the Green's function
is found to dominate (zero temperature, low impurity concentration).
The resistivity saturation phenomenon on the other hand
is expected to appear in the opposite regime where
quasi-particle properties are poorly defined.
In anelastic scattering mechanism (phonons, electron-electron Coulomb
interaction, \ldots)
this regime is usually achieved at large temperature.
A similar role is played in our approach by the bare impurity
scattering rate $\Gamma_0$ which 
rules the broadening of the quasi-particle peak
and the balance between coherent and incoherent contributions.
The resistivity as function of $\Gamma_0$ is shown in Fig. \ref{f-satg}
at half-filling for the uncorrelated case ($U=0$, $Z=1$)
and for an intermediate-highly correlated case ($U/U_c=0.84$, $Z=0.3$).
In low impurity scattering regime, where quasi-particle properties
are well defined and the conductivity is mainly related
to the coherent part, the resistivity increases linearly
with $\Gamma_0$ according Eqs. (\ref{s_coh}) and (\ref{sig0}).
In this framework
the appearing of resistivity saturation
is signalized by the sub-linear behavior pointed out in Fig. \ref{f-satg}.
In the presence of electronic correlation
the magnitude of the resistivity is enhanced, the
sub-linear behavior is more pronounced and it appears for smaller
values of $\Gamma_0$.

As briefly mentioned in the introduction,
a simple criterion for determining the value of the
resistivity saturation $\rho_{\rm sat}$
in non correlated system was presented in Ref. \onlinecite{Gunnarsson_PRB}.
In this analysis the resistivity saturation
is achieved when the spread of the Drude peak
is of the same order of the bandwidth $W$
and the optical conductivity is described by a featureless structure
up to the energy scale $W$. A simple sum rule gives thus
$\rho_{\rm sat}^{\rm SR} \propto W/T_{\rm K}$, where $T_{\rm K}$
is the electronic kinetic energy.
This model was also applied to strongly correlated systems,
as the cuprates.\cite{Gunnarsson_EPL}
In this case, assuming a bandwidth independent
of the degree of correlation,
the scaling of the kinetic energy with
the number of the hole doping $T_{\rm K} \propto \delta$
would predict an increase of the resistivity saturation value
as $\delta \rightarrow 0$, in agreement with the experiments.
However, the assumption of a constant bandwidth $W$ independent
of the electronic correlation is somehow questionable
since the same coherent excitation which are involved in the Drude-like
peak are expected to probe a renormalized dispersion with
the effective bandwidth $W^{\rm eff} = ZW$.
Two different argumentations can be invoked for justifying
both the employing of $W$ and of $W^{\rm eff} = ZW$ in this simple model.
They depend respectively upon two different picture of the physics underlying.
In the first case we assume that the coherent processes are dominant,
and the loss of the Drude peak is due to the broadening of the peak itself.
In this case the bandwidth parameter is consistently given
by $W^{\rm eff} = ZW$ and resistivity saturation is
expected to be achieved when $\Gamma \sim W^{\rm eff}/2$.
In an alternative scenario we can imagine that a narrow Drude-like peak
is still present in the saturation regime, but with a vanishing
spectral weight so that the dominant contribution to the optical
conductivity is given by the incoherent scattering.
A reasonable choice of the appropriate bandwidth parameter
is thus $W$.
The consequences of the different choices are evident by considering
that in the first case the scaling of the effective bandwidth itself
with $\delta$ (or $Z$) would cancel the similar dependence of $T_{\rm K}$
giving a value of $\rho_{\rm sat}$ independent of the electronic correlation,
whereas only in the second case the scaling $\rho_{\rm sat} \sim 1/\delta$
proposed in Ref. \onlinecite{Gunnarsson_EPL} would be operative.

Our model permits us to check on an analytic ground both pictures.
In particular, as representative of the saturation conditions,
we calculate the value of the resistivity
as function of the correlation parameter $Z$
by using Eqs. (\ref{s_coh})-(\ref{s_inc}):
in the first case for $\Gamma=W^{\rm eff}/2$,
which expresses the condition that the broadness of the Drude
peak extends over the entire effective bandwidth,
in the second case for
$\sigma_{\rm coh}=\sigma_{\rm coh-inc}+\sigma_{\rm inc}$,
which determines the crossover at which the incoherent contributions to
the electrical conductivity become of the same order
of the coherent one.
We compare these results with the ones obtained by employing
the SR model of Ref. \onlinecite{Gunnarsson_PRB},
\begin{equation}
\rho_{\rm sat}^{\rm SR} = 
\frac{3V_{\rm cell}}{8 \pi e^2\hbar v_{\rm F}^2 N^2(0)}
\frac{\tilde{W}}{T_{\rm K}},
\label{srmodel}
\end{equation}
with a proper choice of the appropriate bandwidth.
In particular we evaluate
the electron kinetic energy as:
\begin{equation}
T_{\rm K} =
\frac{1}{N_s}\sum_{\bf p} \epsilon_{\bf p}
\int d\omega n_{\rm F}(\omega)
\left[-\frac{1}{\pi}\mbox{Im}G({\bf p},\omega)\right],
\end{equation}
where the quasi-particle impurity scattering rate $\Gamma$
in the Green's function $\mbox{Im}G({\bf p},\omega)$ has been
assumed to be frequency-independent, $\Gamma(\omega)=\Gamma(\omega=0)$,
in agreement with the previous approximation.
It is easy to see that only coherent excitations contribute to $T_{\rm K}$,
so that $T_{\rm K} \simeq Z T_{\rm K}^0$, where
$T_{\rm K}^0$ is the kinetic energy in the absence of correlation.
As mentioned above, more delicate is a proper
definition of bandwidth in both cases.
Following Ref. \onlinecite{Gunnarsson_PRB},
we relate the effective bandwidth to
the lowest order momentum of the dispersion, in particular
in the first case
\begin{equation}
\tilde{W}^{\rm eff} =
\frac{4}{N_s}\sum_{\bf p} |\epsilon_{\bf p}|
\int d\omega 
\left[-\frac{1}{\pi}\mbox{Im}G_{\rm coh}({\bf p},\omega)\right],
\label{wtildeeff}
\end{equation}
and in the second case
\begin{equation}
\tilde{W}^{\rm T} =
\frac{4}{N_s}\sum_{\bf p} |\epsilon_{\bf p}|
\int d\omega 
\left[-\frac{1}{\pi}\mbox{Im}G({\bf p},\omega)\right].
\label{wtilde}
\end{equation}
It is easy to check that the above definitions give
the correct result for the constant DOS model here considered
in the non interacting uncorrelated case, while in the presence
of correlation we recover $\tilde{W}^{\rm T} = W$ and $\tilde{W}^{\rm eff}=ZW$.
A slight modification has been employed with respect to
Ref. \onlinecite{Gunnarsson_PRB}, namely to relate the bandwidth
parameters to the first momentum of the electronic dispersion more than
to the second one, in order to recover the correct
scaling $\tilde{W}^{\rm eff}\propto Z$.

\begin{figure}[t]
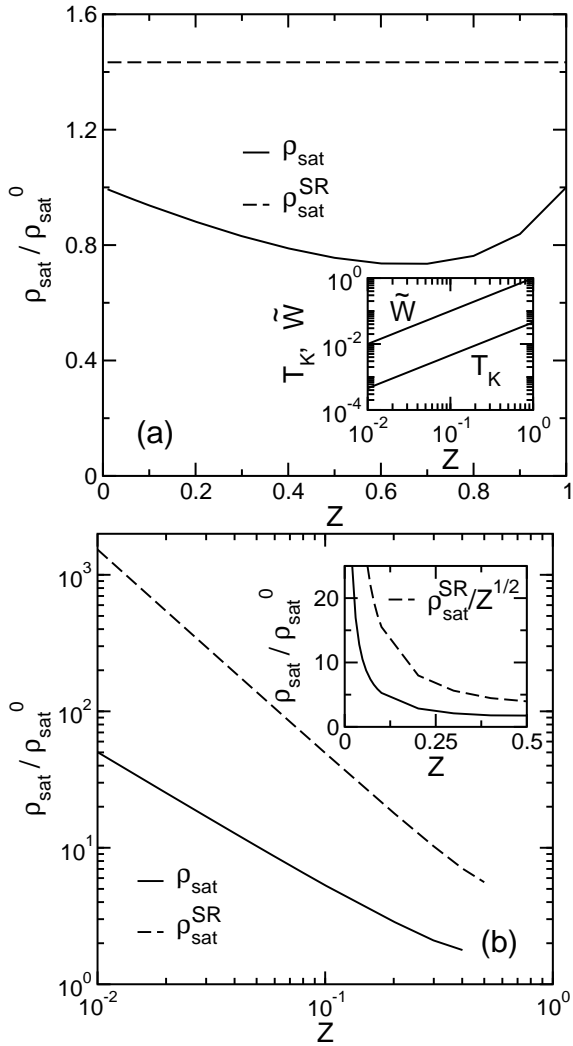

\centerline{
\psfig{figure=Fig6a.eps,width=7.5cm,clip=}}
\centerline{
\psfig{figure=Fig6b.eps,width=7.5cm,clip=}}
\caption{(a) Value of the resistivity saturation as function
of $Z$ evaluated in
explicit way by Eq. (\ref{convol}) (solid line) and by using the
SR model [Eq, (\ref{srmodel}), dashed line] with
the condition $\Gamma = W^{\rm eff}/2$. Inset:linear behavior
of  $T_{\rm K}$ and $\tilde{W} = W^{\rm eff}$ as function of $Z$.
(b) Resistivity saturation limit as function of $Z$, as above,
under the condition
$\sigma_{\rm coh}=\sigma_{\rm coh-inc}+\sigma_{\rm inc}$.
Inset: as in the main picture, on a linear scale,
with $\rho_{\rm sat}^{\rm SR}$ divided by $\sqrt{Z}$ (see text).
}
\label{f-sat}
\end{figure}

In Fig. \ref{f-sat} we compare the direct calculation of the resistivity
saturation value with the SR model in both the cases.
All the resistivity values are expressed in terms
of a ``bare'' resistivity saturation value,
\begin{equation}
\rho_{\rm sat}^0
= \frac{3 V_{\rm cell}}{\hbar e^2v_{\rm F}^2 N^2(0)[1+2/\pi]},
\end{equation}
which represents the saturation value expected for
an infinite bandwidth uncorrelated system, and which depends
only on structural and band-structure properties.
As discussed above, the criterion for the resistivity
saturation has been determined by $\Gamma=W^{\rm eff}/2$
in the first case (panel a) and by
$\sigma_{\rm coh}=\sigma_{\rm coh-inc}+\sigma_{\rm inc}$
in the second case (panel b).
The agreement between the direct calculation of the electrical resistivity
by means of the Kubo's formula Eqs. (\ref{convol}) and the SR model
is quite good considering the simplicity of the SR model.

Panel (a) describes the case where the coherent part of the electrical
conductivity is predominant and the
resistivity saturation is driven
by the broadening of the Drude-like peak.
The disappearing of the spectral weight $Z$ in this case is accompanied
by a similar scaling of $T_{\rm K}, \tilde{W}^{\rm eff} \propto Z$
(see inset).
These two similar dependences cancel each other both
in the SR model and in the coherent contribution of Eq. (\ref{convol}),
yielding a resistivity saturation value independent of the correlation degree.
The weak dependence of $\rho_{\rm sat}$ on $Z$
in the direct evaluation of the resistivity saturation
by using  Eq. (\ref{convol}) is indeed
due to the small incoherent contributions to the
transport properties and it would not be present if only
the coherent part would be taken into account.

The alternative scenario for saturation is shown in Fig. \ref{f-sat}b
where the criterion for resistivity saturation is related
to the predominance of the incoherent parts of the conductivity.
In this case the vanishing amount of coherent states, evaluated by
$T_{\rm K}$ ($T_{\rm K} \rightarrow 0$ for $Z \rightarrow 0$),
is not compensate by a corresponding reduction of the effective
bandwidth $\tilde{W}^{\rm T}$, and the resistivity saturation value
$\rho_{\rm sat}$ is expected to diverge as the correlation effects
increase $Z \rightarrow 0$. Once more, these simple physical argumentations,
employed in the simple SR model, give a qualitative insight which agrees
in a satisfactory way, with the explicit calculation of the
resistivity through the Kubo's formula, Eq. (\ref{convol}),
although the scaling law is different. In particular
the explicit calculation of $\rho$ from Eq. (\ref{convol})
predicts an inversely proportional dependence of $\rho_{\rm sat}$
as function of the spectral weight $Z$, $\rho_{\rm sat} \propto 1/Z$,
whereas $\rho_{\rm sat}^{\rm SR} \propto 1/Z^{3/2}$.
This discrepancy is related to the somehow anomalous
behavior of $T_{\rm K}$, $T_{\rm K} \propto Z^{3/2}$,
which stems from the fact that, in the strongly correlated
limit $Z \rightarrow 0$, $W^{\rm eff} \ll \Gamma $
($W^{\rm eff} \propto Z$, $\Gamma \propto \sqrt{Z}$)
and $T_{\rm K} \propto W^{\rm eff}\Gamma$.
In the inset of Fig. \ref{f-sat}b we compare thus the explicit
evaluation of $\rho_{\rm sat}$ from Eq. (\ref{convol}) with
$\rho_{\rm sat}^{\rm SR}/\sqrt{Z}$, showing indeed
a good agreement between the two quantities.

Summarizing the present results, we have investigated the dependence
of the resistivity saturation in two alternative scenarios.
In the first case the resistivity saturation is achieved when the
coherent Drude-like peak broadens over the coherent bandwidth,
ruled by the condition $\Gamma \simeq ZW/2$.
In the second case the resistivity saturation is expected
to appear when the incoherent contributions
to the electrical conductivity become dominant with
respect to the coherent one, and it is related to the
regime $\sigma_{\rm coh} \simeq \sigma_{\rm coh-inc}+\sigma_{\rm inc}$.
We have shown that in the first scenario the resistivity saturation value
is essentially independent of the degree of electronic correlation,
and it could not account for the experimental lack of saturation
in the underdoped region of cuprates.
On the other hand, the resistivity saturation is expected to scale
in the second scenario
with the reduced spectral weight $Z$ due to the correlation effects,
$\rho_{\rm sat} \propto 1/Z$.
This result is in a good agreement with recent experimental
data by Takenaka {et al.},\cite{Takenaka} which nicely show the
$\rho_{\rm sat} \propto 1/Z$ behaviour of the DC
resistivity
in the low doping region $0 < x < 0.3$ where
the spectral weight $Z$ itself is expected to scale with $x$,
$Z \propto x$.\cite{Yoshida}
Similar results were also found by
numerical calculation
based on exact diagonalization technique.\cite{Gunnarsson_EPL}

\begin{figure}[t]
\centerline{
\psfig{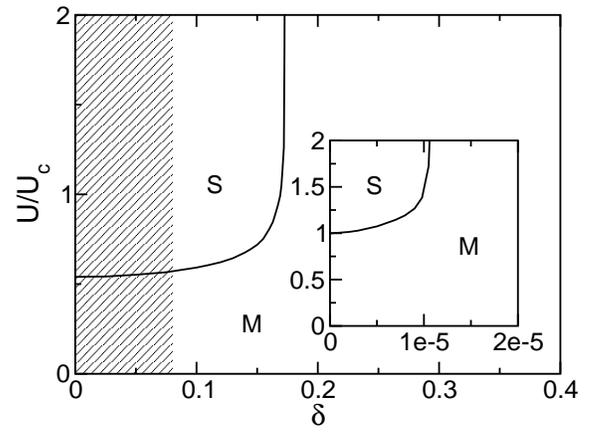}}
\caption{Phase diagram for the saturation effects
in the $U$ vs. $\delta$ space. The label S indicates regions where
saturation occurs according the condition
$\sigma_{\rm coh}<\sigma_{\rm coh-inc}+\sigma_{\rm inc}$,
while in the M region the system presents Drude-like
metallic behaviour.
Main frame: $\Gamma_0/W = 1.0$; inset:
$\Gamma_0/W = 0.05$. In both cases $\pi N(0) V_{\rm imp}=0.005$.
}
\label{f-phase}
\end{figure}

The resulting phase diagram in the $U$-$\delta$ space
for the relevance of the saturation effects is shown in Fig. \ref{f-phase}.
According the above discussion the criterion for saturation as
been determined by the condition 
$\sigma_{\rm coh} = \sigma_{\rm coh-inc}+\sigma_{\rm inc}$.
Here the label M indicates region where the conductivity is
dominated by the coherent part and the system behaves like a strongly
correlated metal but with metallic characteristics, while
in the S region the incoherent contributions prevail
and the saturation behaviour is expected.
Since the coherent and incoherent contributions to the conductivity
scale in a different way with the bare impurity scattering rate
$\sigma_{\rm coh} \propto 1/\Gamma_0$,
$\sigma_{\rm coh-inc}, \sigma_{\rm inc} \sim$ const., the balance
is strongly dependent on $\Gamma_0$.
In the main frame of Fig. \ref{f-phase} we show the phase diagram
for a sizable scattering rate $\Gamma_0/W = 0.1$, where saturation effects
extend up to $\delta \sim 0.175$ in the correlated regime.
The shadow area points out qualitatively the region where Van Hove
singularity effects are expected in the two-dimensional tight-binding
model appropriate for cuprates and where the constant DOS model
is expected to fail.
The inset shows also the very weakly impurity case $\Gamma_0/W = 0.05$.
In this case the coherent part is almost always dominant
and strong correlation effects, for $U \ge U_c$ and $\delta \sim 0$
where $Z \rightarrow 0$ are needed in order to show saturation.
We remind that in the generalization of the present
analysis to the electron-phonon scattering mechanism, the
scattering rate $\Gamma_0$ plays the role of the temperature
dependent scattering rate, which can be easily of the order
of the bandwidth $\Gamma_0^{\rm el-ph}(T)/W$ for large enough temperatures.

\section{Conclusion}

Aim of this work has been to investigate
electronic correlation effects on the quasi-particle and transport
properties. A key role in this context is thought to be played
by the reduction of the coherent spectral weight accompanied
by the onset of localized state.
For instance the lack of resistivity saturation at the Mott-Ioffe-Regel
limit in cuprates superconductors has been recently
related to an effective renormalization of single-particle quantities, as
the kinetic energy, due to the strong electronic correlation.
Although some analytical models based on a sum rule conservation
have been proposed, the effects of the electronic correlation
have been mainly studied by means of numerical approaches.

In this paper we have introduced a simple phenomenological model
in order to describe in an analytical way the transfer of
spectral weight from coherent to incoherent states as the degree
of electronic correlation is increased.
We have studied the interplay between electronic correlation and
impurity scattering in quasi-particle and transport properties.
We identify two different regimes,
a low correlated regime, where the effective
bandwidth of coherent states is much greater than the impurity
scattering rate, and a highly correlated regime where the renormalization
effects induced by the electronic correlation yield an effective
coherent bandwidth $W^{\rm eff}$ smaller than
the impurity scattering rate itself.
Contrary to what one should expect, correlation effects in this latter
regime increase the quasi-particle life-time.

We have also evaluated the electrical resistivity by means
of the current-current response function.
Within the simple bubble approximation, we have shown that correlation
effects enhance the resistivity $\rho$ (suppress the conductivity $\sigma$)
according the scaling law $\rho \propto 1/Z$
in both low and highly correlated regimes.
This results suggest that the `'phenomenological''
quasi-particle and transport scattering rate,
respectively $\Gamma$ and $\Gamma_{\rm tr}$, as extracted by the experiment
within a quasi-particle analysis, should scale in an opposite
way upon the relevance of the electronic correlation.
Our analysis permits in addition to test in a direct way
some predictions of a simple sum-rule model employed
to describe the resistivity saturation limit.
We found that the lack of resistivity saturation
in cuprates can be interpreted as an effective enhancement
of the saturation limit as correlation effects are increased approaching
the half-filling case, in agreement with
Refs \onlinecite{Gunnarsson_EPL}. This regime is however
achieved when the incoherent contributions to the electrical
conductivity become of the same order to the coherent one.

This work was partially funded by
the MIUR projects FIRB RBAU017S8R and COFIN 2003.
We would like also to thank M.L. Kuli\'c and R. Zeyher for having
pointed out a mistake in a previous version of the manuscript.

\appendix

\section{Analytical expressions for some
coherent and incoherent one-particle and two-particle quantities}
\label{appendice} 

In this appendix we summarize and provide 
the analytical expressions for some one- and two-particle quantities
which have been employed in the previous
sections.

We first consider the impurity self-energy defined in
Eqs. (\ref{gammadelta})-(\ref{gammaz}), which are expressed as function
of the real and imaginary part of the local electron Green's function,
respectively $G'_{\rm loc}\equiv \mbox{Re}G_{\rm loc}(\omega=0)$ 
and $G''_{\rm loc}\equiv \mbox{Im}G_{\rm loc}(\omega=0)$.
Both of them are characterized by a coherent and an incoherent contribution.
Simple analytical expressions for these quantities
can be obtain from Eqs. (\ref{Greenc})-(\ref{Green1}):
\begin{eqnarray}
G'_{\rm loc}
&=&
R_{\rm coh}+R_{\rm inc},
\\
G''_{\rm loc}
&=&
I_{\rm coh}+I_{\rm inc},
\end{eqnarray}
where
\begin{eqnarray}
R_{\rm coh}
&=&
\frac{1}{2}\log \left[
\frac{(ZW/2-\mu)^2+\Gamma^2}{(ZW/2+\mu)^2+\Gamma^2} 
\right],
\label{g_{coh}}
\\
R_{\rm inc}
&=&
\frac{1}{2}
\left\{ \log \left[
\frac{[(1-n/2)W/2-\mu+U/2]^2+\Gamma^2}{[(1-n/2)W/2+\mu-U/2]^2+\Gamma^2} 
\right]  \right.
\nonumber\\
&&
+
\left.
\log \left[
\frac{(nW/4-\mu-U/2)^2+\Gamma^2}{(nW/4+\mu+U/2)^2+\Gamma^2} 
\right]
\right\},
\label{g_{inc}}
\\
I_{\rm coh}
&=&
\arctan\left[\frac{ZW/2-\mu }{\Gamma}\right]
+\arctan\left[\frac{ZW/2+\mu}{\Gamma}\right],
\label{autoconsistente} 
\\
I_{\rm inc}
&=&
\left\{
\arctan \left[ \frac{(1-n/2)W/2-\mu+U/2}
{\Gamma}\right]
\right.
\nonumber\\
&&
+\arctan \left[\frac{(1-n/2)W/2+\mu-U/2}{\Gamma}\right]
\nonumber\\
&&
+\arctan \left[\frac{nW/4-\mu-U/2}{\Gamma}\right]
\nonumber\\
&&
+\left.
\arctan \left[\frac{nW/4+\mu+U/2}{\Gamma}\right]
\right\}.
\label{f_{inc}}
\end{eqnarray}

Some of the same quantities are employed also in the definition of the
electrical conductivity
as well as in the transport impurity scattering rate
evaluated in the simple bubble approximation,
respectively Eqs. (\ref{s_coh})-(\ref{s_inc}) and
(\ref{Gamma_transport}), together with the quantity
\begin{equation}
A=\Gamma 
\left[ 
\frac{ZW/2-\mu}{(ZW/2-\mu)^2+\Gamma^2}
+\frac{ZW/2+\mu}{(ZW/2+\mu)^2+\Gamma^2}
\right].
\end{equation}


\begin{thebibliography}{100}
\bibitem{notegreen}
Since in this paper we mainly use only retarded Green's functions,
unless explicitly stated,
the frequency $\omega$ is understood to contain an infinitesimally
small imaginary part $\omega \equiv \omega +i0^+$. Such infinitesimally
small imaginary part of course does not play any role when a finite
imaginary self-energy is present.
\end{thebibliography}
\end{document}